# Electrical control of optical orientation of neutral and negatively charged excitons in n-type semiconductor quantum well

R.I. Dzhioev, V.L. Korenev[*], M.V. Lazarev, V.F. Sapega

A.  F. Ioffe Physical Technical Institute, St. Petersburg, 194021 Russia

D. Gammon, A.S. Bracker

*Naval Research Laboratory, Washington DC 20375, USA*

*We report a giant electric field induced increase of spin orientation of excitons in n-type GaAs/AlGaAs quantum well. It correlates strongly with the formation of negatively charged excitons (trions) in the photoluminescence spectra. Under resonant excitation of neutral heavy-hole excitons, the polarization of excitons and trions increases dramatically with electrical injection of electrons within the narrow exciton-trion bias transition in the PL spectra, implying a polarization sensitivity of 200 % per Volt. This effect results from a very efficient trapping of neutral excitons by the quantum well interfacial fluctuations (so-called "natural" quantum dots) containing resident electrons.*

PACS: 78.67.–n, 73.21.–b, 85.75.–d

*Corresponding author: Korenev@orient.ioffe.ru*

Electrical manipulation of the spin degree of freedom has been attracting considerable interest [1] due to its potential applications. It was observed [2,3] that under optical orientation conditions the photoluminescence (PL) circular polarization in n-type quantum wells is very large at high electron concentration (higher than critical density n>2*10$^{10}$ cm$^{-2}$) resulting from the screening-induced quenching of neutral excitons and spin relaxation due to the long-range electron-hole exchange interaction [4]. At lower concentrations the neutral excitons are not quenched but tend at low temperature to form the negatively charged exciton (trion) [5]. Optical orientation of trions in nanostructures has been done first in quantum-sized InP islands [6], followed by a number of publications [7, 8] in various heterostructures. Although only the hole can be polarized in the trion ground state (electrons form singlet), its spin polarization is determined by the spin orientation of neutral excitons and resident electrons at the moment of trion formation [6]. Single dot spectroscopy [8] revealed that the optical orientation of single trion PL varies strongly with n-type charging produced by an externally applied bias. Magnetic depolarization measurements in Voigt geometry (the Hanle effect [9]) relates this change mainly with the optical pumping of resident electrons whose polarization is decreased substantially due to electrical injection of non-polarized electrons.

Here we show a striking correlation between the exciton-trion intensity ratio and PL circular polarization at different biases in the ensemble PL spectra of a 4 nm GaAs/AlGaAs quantum well (QW). Namely, the onset of the trion peak in the PL spectra correlates with a sharp increase of zero-field PL circular polarization from 10 % to 70 %. The Hanle effect enables us to separate unambiguously the contribution of resident electrons to the PL circular polarization from the contribution of neutral excitons. Although the orientation of resident electrons decreases with the electrical injection of electrons in accord with [8], the optical orientation of neutral excitons that form trions is found to increase dramatically (from 2 % up to 70 %) within the narrow bias range (0.3 V only). It means enormous spin polarization sensitivity to electric field ~200 %/V. We explain this effect as a result of competition between spin relaxation of neutral excitons and the trapping process, with the latter becoming very efficient in the presence of resident electrons on the localization sites. The importance of the trapping dynamics in optical orientation was not previously considered, but as we show here it can be very important for localized systems such as quantum dots.

The 4nm wide GaAs/AlGaAs quantum well is contained within a Schottky diode that provides electron injection with an applied bias [8, 10]. The sample was placed in a liquid-helium cryostat and pumped quasi-resonantly by a tunable Ti-sapphire laser, with the circular polarization of light being alternated in sign at a frequency of 26.61 kHz with a photoelastic quartz modulator. This permitted us to eliminate the effect of the lattice nuclear polarization on the optical orientation of the electrons. The PL polarization was measured in the reflection geometry by a circular-polarization analyzer. The electronics provided measurement of the *effective* degree of circular polarization $\rho = (I_+^+ - I_+^-)/(I_+^+ + I_+^-)$, where $I_+^+$, $I_+^-$ are the intensities of the $\sigma^+$ PL component under the $\sigma^+$ and $\sigma^-$ pumping, respectively. In our case, ρ may be considered as a Stokes parameter characterizing the PL circular polarization because the circular dichroism effects are insignificant [11]. The magnetic field was applied in Faraday or in Voigt geometries with the use of a superconductive split-coil magnet.

The heavy-hole neutral exciton state in [001]-grown quantum well consists of four spin sublevels, taking into account electron spin up $(\uparrow)$ and down $(\downarrow)$ as well as heavy-hole angular momentum projections $m_h$=+3/2 ($\Uparrow$) and -3/2 ($\Downarrow$). Excitonic state, bright exciton, $|\downarrow\Uparrow\rangle$ $(|\uparrow\Downarrow\rangle)$ with total momentum +1 (-1) is optically allowed in $\sigma^+$ ($\sigma^-$) polarization whereas the dark exciton state $|\uparrow\Uparrow\rangle$ $(|\downarrow\Downarrow\rangle)$ with parallel electron and hole spins is optically forbidden. Near resonance (laser detuning energy is smaller than exciton binding energy) the right-handed ($\sigma^+$) circularly polarized light at normal incidence creates in the quantum well bright excitons (X) with angular momentum projection +1 along growth direction. Spin relaxation of electron, hole or mutual electron-hole spin flips distribute excitons over all four states, reducing the initial optical orientation [12]. Capture of the excitons by the monolayer-width fluctuations of the QW interface (so-called "interfacial" or "natural" quantum dots [13]) leads to the Stokes shifted PL. Exciton luminescence (X-line) originates from excitons localized in empty dots. The degree of circular polarization of the X-line is equal to the polarization $\rho_X$ of bright excitons. At low temperature resident electrons populate (bias-dependent) part of the dots. Trapping of neutral excitons by filled dots creates negatively charged excitons – trions whose luminescence produces trion PL line (T-line). In the trion ground state the two electrons form a singlet. The total angular momentum of the trion states $|+3/2\rangle = |\uparrow\downarrow\Uparrow\rangle$ and $|-3/2\rangle = |\uparrow\downarrow\Downarrow\rangle$ is determined by the momentum of the hole. The degree of circular polarization of the T-line $\rho_T$ is equal to the hole polarization, with the latter being determined by the polarizations of neutral excitons (both bright and dark) and resident electrons at the moment of trion formation [6].

Solid lines in Fig. 1 shows the bias dependence of PL intensity (left axis of each graph) spectra at quasi-resonant excitation ($\lambda$=7555 A) of the 4 nm GaAs/AlGaAs QW at 5 K. One can see two pairs of lines. Within each pair the high-energy line shows recombination of neutral excitons (X) whereas the low-energy line corresponds to the recombination of negatively charged excitons – trions (T) [14]. These transitions come from recombination of quasiparticles localized on monolayer width islands (dots) as proven by single-dot spectroscopy [13]. Each pair of lines belongs to two QW widths differing by 1 monolayer – upper monolayer (UML, 15 monolayers in width) and lower monolayer (LML, 16 monolayers in width) respectively. One can see that increase of bias from –1.1 V to –0.8 V favors trion PL at the expense of exciton PL due to the filling of the QW with electrons.

Polarization spectra (right axis of each panel) are measured in a B=0 T (dotted) and B=5 T (dashed) magnetic field applied in Faraday geometry (PL intensity increases with B slightly and is not shown here). Figure 1 shows that application of magnetic field increases strongly the polarization $\rho_X$ of bright excitons. It is well known that the optical orientation of excitons localized on anisotropic islands is suppressed as a result of electron-hole (e-h) anisotropic exchange interaction mixing states with momentum projection +1 and −1 [13, 15]. The magnetic field cancels the effect of exchange and restores the optical orientation of localized excitons. This behavior is clearly seen for the X-line originating from the recombination of bright excitons localized on "natural" quantum dots [8]. In turn, the magnetic field has *only a weak effect* on the trion polarization (change of $\rho_T$ at $V_g$=-1.10 V is due to the overlap with two excitonic transitions). This important result shows that the trion is formed by trapping of a free exciton (which is not depolarized by anisotropic exchange interaction) into a dot with an electron. The absence of an "anisotropic exchange history" of trion rules out the alternative channel of trion formation consisting in the initial localization of the exciton in an empty dot followed by the electron tunneling from the doped substrate [16].

Reversed bias $V_g$=-1.10…-1.00 V depletes the well of electrons, thus favoring excitonic PL. The restored circular polarization of bright excitons $\rho_X(5T)$ is about 50 % for UML and does not reach the theoretical limit 100 % under resonant excitation. This may be the result of long-range electron-hole exchange interaction [4] mixing bright states and depolarizing free excitons. In turn, $\rho_X(5T)$ is larger than the polarization of trions $\rho_T(5T)$ (about 20 % for UML at $V_g$=-1.00 V) in this bias range. This suggests efficient single-hole spin flip processes within the exciton *before* trion formation [17]. The phonon assisted hole spin-flip transfers the optically created bright $|\downarrow\Uparrow\rangle$ exciton into the $|\downarrow\Downarrow\rangle$ dark one. The latter does not affect the polarization of the X-line but decreases substantially the polarization of the T-line via the formation of $|\uparrow\downarrow\Downarrow\rangle$ trions with total angular momentum –3/2 (sometimes it leads to negative polarization, see Fig.1 for LML and Ref. [8]). Thus our results at large reversed bias suggest efficient spin relaxation of free excitons before localization.

Filling the QW with electrons ($V_g$>-1.00 V) makes the T-line stronger than the X-line and increases the circular polarization of the PL. At $V_g$=-0.80 V the circular polarization of T-line reaches 70 %. Although the excitonic PL is much weaker, one can see the effect of magnetic field on PL polarization at the X-line energy. It means that part of the dots is empty of electrons even at this bias, i.e. all resident electrons are able to find localization sites.

The bias dependence of PL polarization on Fig.1 suggests an enhancement of the optical orientation of excitons when in the presence of resident electrons in the dots. The Hanle effect measurements (Voigt geometry) support this conclusion. The Hanle effect separates the contributions of excitons and resident electrons to the circular polarization of the trion PL. Larmor precession of trions is absent because the in-plane g-factor of holes is very close to zero in [001]-GaAs/AlGaAs quantum wells [18]. Axially symmetric electron-hole exchange interaction splits bright and dark exciton states [12], preventing electron spin precession within the exciton. In contrast, the spins of unbound electrons undergo Larmor precession. Figure 2 shows the Hanle effect measured on exciton (a) and trion (b) lines at $V_g$>-1 V. One can see incomplete depolarization: the initial change is followed by saturation at a level $\rho_{sat}$ strongly depending on bias. Close inspection of initial part shows that it is different for X and T lines. The polarization of X-line decreases with magnetic field. In contrast, polarization of the T-line grows initially at low fields followed by a slow decay similar to that for X-line. This "M-letter" shape of the T-line with a small inverted peak at low fields was observed previously in wide (14 nm) well [19]. The narrow Hanle peak was explained as Larmor precession of localized resident electron spin with spin lifetime falling into nanosecond range. The Hanle effect in single dot located in the 3 nm QW also revealed the narrow peak due to the optical pumping of resident electrons [8]. It was found that the contribution of resident electrons to the polarization decreases with injection of non-polarized electrons in agreement with our results: the narrow peak disappears at higher bias. The nature of the wide peak is less clear. It was interpreted as a rotation of electron spin within the exciton [19] (see also Ref. [20]), giving shorter spin relaxation time (~100 ps). Although this is reasonable for the wide (14 nm) QW with reduced value of exchange splitting of bright and dark states [21], it seems unlikely for the narrow QW under study: a magnetic field B~60 kG is required to mix the bright and dark states [18].

Nevertheless Fig.2a shows that the halfwidth of the wide peak is about 2 kG and varies slowly with bias, suggesting a photoexcitation origin. We leave this problem for future studies.

Here we concentrate on the much more striking effect - the steep increase of the saturation value $\rho_{sat}$ with gate voltage for the T-line [22]. This unambiguously points to the enhancement of the optical orientation of excitons captured by the charged quantum dots. Indeed, resident electrons are depolarized by magnetic field and do not contribute to the polarization of the trion. In turn, the spin relaxation of localized trions is blocked [17]. Therefore polarization of T-line $\rho_{sat}$ (Fig.3b) reflects the polarization of neutral excitons at the moment of trion formation. It is known [9] that non-equilibrium polarization is determined by the competition between lifetime $\tau$ and spin relaxation time $\tau_s$. In our case $\tau$ is the lifetime of exciton before localization. An increase of exciton polarization with electron concentration implies an increase of the ratio $\tau_s/\tau$. There are two possibilities:

(1) Spin relaxation time of excitons increases with density of localized electrons. This seems unlikely because (i) in the absence of free carriers screening is negligible; (ii) the scattering of excitons by non-polarized resident electrons should induce only additional depolarization of excitons; (iii) one should explain the simultaneous suppression of exciton spin relaxation mechanisms due to long-range exchange and single hole spin flip, whose origins are very different. For these reasons we rule out the quenching of spin relaxation of excitons.

(2) Exciton lifetime $\tau$ decreases. The experimental data in Fig.3 support this conclusion. The filled circles in Fig.3 show the relative strength $\eta = I_T/(I_T + I_X)$ of T-line, where $I_T (I_{X^0})$ are the intensities of trion (exciton) PL. Open symbols show the bias dependence of saturated circular polarization degree $\rho_{sat}$ obtained from Fig.2(b). The degree $\rho_{sat}$ shows that the steepest rise from 2 % (at -1.1 V) up to 70 % (at -0.8 V) occurs within a bias range of 0.3 V, implying sensitivity of about 200 %/V. One can see a *striking correlation between the ratio η of trion/exciton PL intensities and the optical orientation of trions $\rho_{sat}$*. This means that the previous interpretations [2, 3] that did not consider the trion formation process fail to explain the data. The observed correlation implies a significant shortening of the exciton capture time [23]: it both increases the optical orientation of excitons that form trions and favors the trion peak. At low temperature resident electrons are localized on the interfacial quantum dots [24]. Thus trions are formed by capturing of excitons into those dots filled with single electrons [25]. The difference between the cases of an empty QW or a QW filled with electrons is that the photo-excited excitons are trapped by empty dots or by dots filled with one electron, respectively. Thus we must conclude that *trapping of excitons by populated dots is accelerated strongly in comparison with empty dots*.

Consider a pedagogical model illustrating the final conclusion. Non-polarized resident electrons fill the interfacial dots with the filling factor f (between 0 and 1) depending on bias. Quasi-resonant $\sigma^+$–light creates excitons in $|+1\rangle$ state. For simplicity we assume that the main spin relaxation mechanism is due to the long-range interaction between [4] electron and hole in the exciton that mixes states $\pm 1$ with time $\tau_s$. Thus we consider only bright states. In the absence of free electrons the screening effect is negligible and we can neglect the dependence of $\tau_s$ on filling factor f. In turn, the exciton lifetime $\tau$ before localization can be written phenomenologically as

$$\frac{1}{\tau} = \frac{1-f}{\tau_e} + \frac{f}{\tau_f} \qquad (1)$$

where $\tau_e$ is exciton lifetime for empty case (f=0) and $\tau_f$ is that for the filled (f=1) case. The steady-state polarization of excitons is given by usual formula [9]

$$P_X = \frac{\tau_s}{\tau_s + \tau} \qquad (2)$$

Exciton capture by a dot with one electron forms a trion whose polarization $P_T$ is equal to $P_X$, given by Eq.(2). Spin relaxation of trions can be neglected as we discussed above [17]. As a result, the PL polarization (both trion' and exciton') is determined by Eq.(2). The intensities of the exciton ($I_X$) and trion ($I_T$) recombination are proportional to the capture rates, $\frac{1-f}{\tau_e}$ and $\frac{f}{\tau_f}$, respectively. As a result, the relative trion intensity is given by

$$\eta = \frac{I_T}{I_T + I_X} = f\frac{\tau}{\tau_f} \quad (3)$$

Equations (1-3) show that the lifetime difference causes the correlation between $P_X$ and $\eta$. If $\tau_e$ and $\tau_f$ were the same, the exciton lifetime $\tau=\tau_e=\tau_f$ (and polarization $P_X$) would not depend on filling factor f (bias) whereas the trion contribution to the total PL increases linearly with f: $\eta=f$. The solid (dashed) curve on Fig.4 is the result of calculation using Eqs.(1-3) with $\tau_f = \tau_s$, $\tau_e = 10\tau_f$. In other words the bright exciton is doing a lot of spin flips before localization in the empty case ($\tau_s \ll \tau_e$) whereas it has no time to be flipped in the filled case ($\tau_s = \tau_f$). The reason for such a giant difference may come from the change of trapping mechanism: it is phonon-assisted trapping into the empty dot and Auger-like into the filled dot. Inset on Fig.4 illustrates this point. Bright exciton in $|+1\rangle$ state scatters with resident quantum dot electron having spin antiparallel to the electron spin in exciton. This creates $|+3/2\rangle$ trion within the dot on an excited state followed by its spin-conserving relaxation to the ground state [26]. This process cancels out the electron-hole anisotropic exchange that decreases the degree of circular polarization of PL (two electrons form singlet with zero spin [27]). It does not require phonon assistance, thus accelerating trapping of excitons (in comparison to the case of empty dots) and preserving their spin orientation. This qualitatively explains the data.

In conclusion, we report a striking correlation between the exciton-trion intensity ratio and the optical orientation of both neutral and negatively charged excitons at different biases in the ensemble PL spectra of a GaAs/AlGaAs quantum well. The optical orientation of excitons is found to increase dramatically within the narrow bias range, showing an enormous polarization sensitivity of ~200 %/V. We explain this effect as a result of very efficient trapping of neutral excitons by islands containing the resident electron.

This work was supported by NSA/ARO, RFBR, RSSF, ONR.

Figure Captions

Figure 1. Photoluminescence spectra of intensity and degree of circular polarization measured in a 4 nm GaAs/AlGaAs quantum well at different biases. Solid lines show PL intensity spectra, where X and T label exciton and trion PL peaks. Polarization spectrum at B=0 (B=5 T, Faraday geometry) is shown by dotted (dashed) line. UML (LML) denotes upper (lower) monolayer quantum well of 15 ML (16 ML) width.

Figure 2. The Hanle effect data (magnetic field is applied in Voigt geometry) at different biases. (a) neutral exciton (b) negatively charged exciton (trion)

Figure 3. Bias dependence of saturation polarization $\rho_{sat}$ (open circles) as deduced from the Hanle data for the trion line. Filled circles show the trion contribution $\eta = I_T/(I_T + I_X)$ to the total PL intensity.

Figure 4. Filling factor dependence of the polarization of bright excitons (solid curve) and the parameter $\eta$ (dashed curve), calculated with $\tau_e=10\tau_f$ and $\tau_s=\tau_f$. Inset shows a cartoon of the Auger-assisted formation of a trion localized in a "natural" quantum dot.

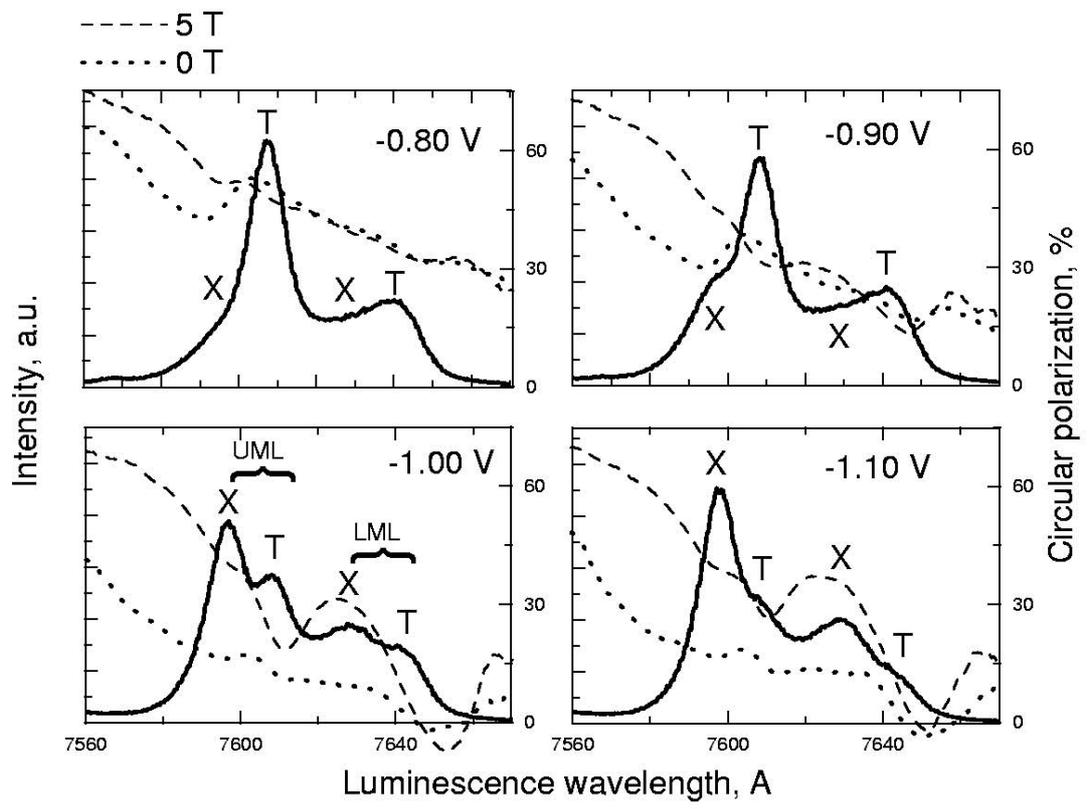

Figure 1

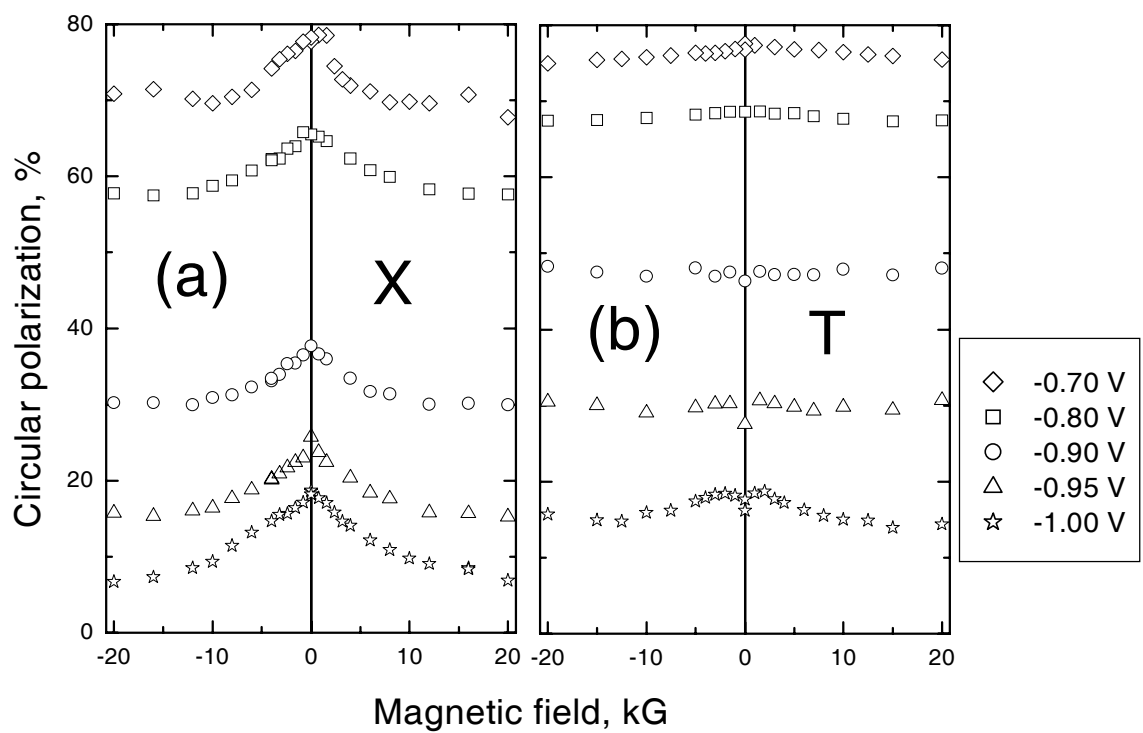

Figure 2

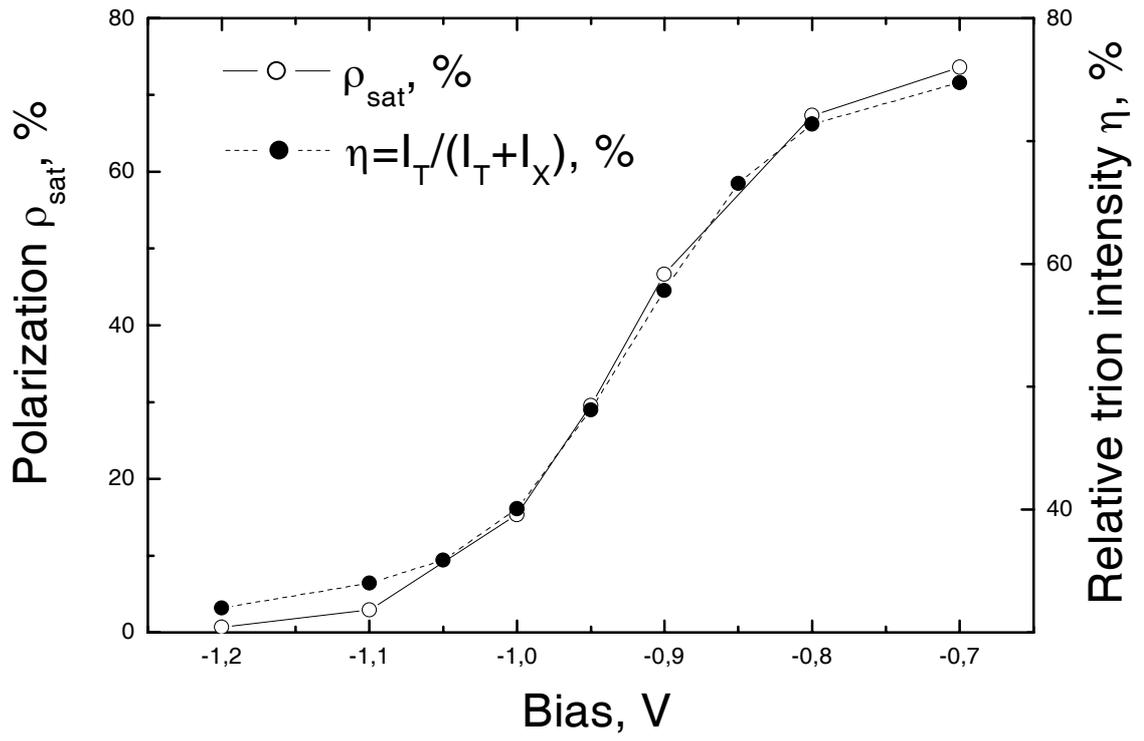

Figure 3

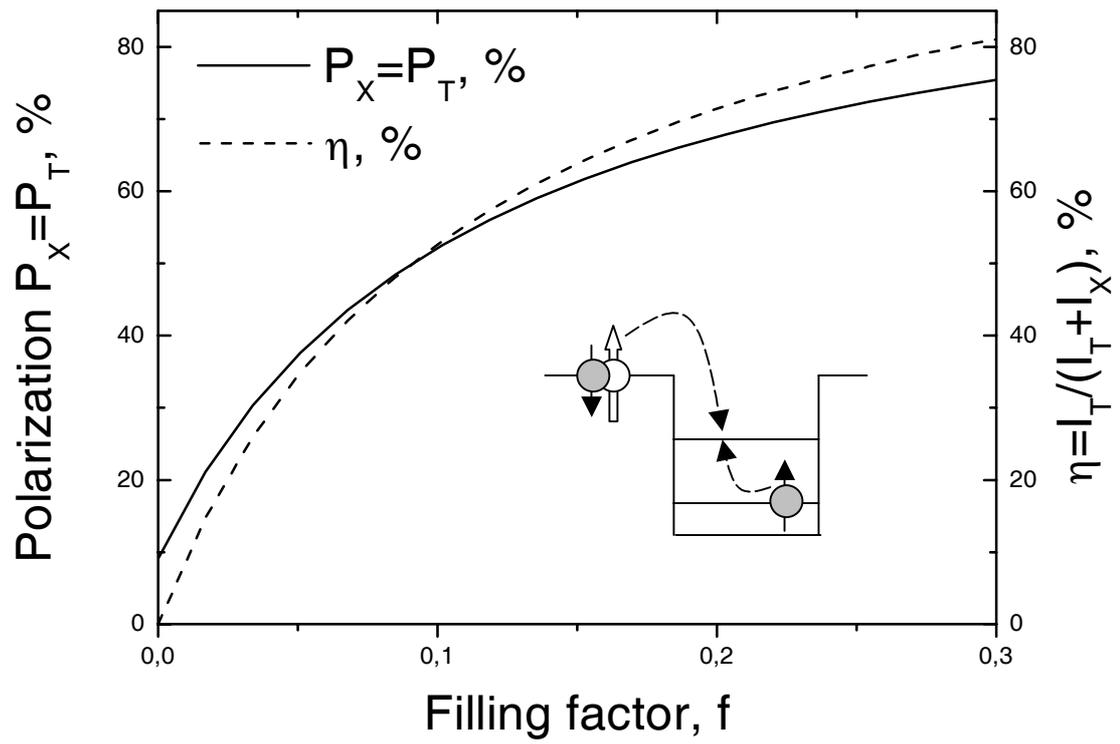

Figure 4